\documentclass[prb,twocolumn,showpacs,floatfix,amsmath,amssymb,superscriptaddress]{revtex4-1}
\usepackage{amsfonts}
\usepackage{stmaryrd}
\usepackage{bbm}
\usepackage{mathrsfs}
\usepackage{tipa}
\usepackage{amssymb}
\usepackage{txfonts}
\usepackage{graphicx}
\usepackage{dcolumn}
\usepackage{epstopdf}
\usepackage[colorlinks,linkcolor=blue,urlcolor=blue,citecolor=blue]{hyperref}
\usepackage{multirow}
\usepackage{subfigure}
\usepackage{url}
\usepackage{upgreek}
\usepackage[utf8]{inputenc}
\usepackage[english]{babel}

\begin{document}

\newcommand*{\cm}{cm$^{-1}$\,}
\newcommand*{\Tc}{T$_c$\,}

\title{Experimental Observation of Quantum Many-Body Excitations of $E_8$ Symmetry in the Ising Chain Ferromagnet CoNb$_2$O$_6$}

\author{Kirill Amelin}
\affiliation{National Institute of Chemical Physics and Biophysics, 12618 Tallinn, Estonia}

\author{Johannes Engelmayer}
\affiliation{Institute of Physics II, University of Cologne, 50937 Cologne, Germany}

\author{Johan Viirok}
\author{Urmas Nagel}
\author{Toomas Rõõm}
\affiliation{National Institute of Chemical Physics and Biophysics, 12618 Tallinn, Estonia}

\author{Thomas Lorenz}
\author{Zhe Wang}
\email{zhewang@ph2.uni-koeln.de}
\affiliation{Institute of Physics II, University of Cologne, 50937 Cologne, Germany}

\date{\today}

\begin{abstract}
Close to the quantum critical point of the transverse-field Ising spin-chain model, an exotic dynamic spectrum was predicted to emerge upon a perturbative longitudinal field. The dynamic spectrum consists of eight particles and is governed by the symmetry of the $E_8$ Lie algebra. Here we report on high-resolution terahertz spectroscopy of quantum spin dynamics in the ferromagnetic Ising-chain material CoNb$_2$O$_6$. At 0.25~K in the magnetically ordered phase we identify characteristics of the first six $E_8$ particles, \textbf{m}$_1$ to \textbf{m}$_6$, and the two-particle (\textbf{m}$_1+$\textbf{m}$_2$) continuum in an applied transverse magnetic field of $B_c^{1D}=4.75$~T, before the three-dimensional magnetic order is suppressed above $B_c^{3D}\approx 5.3$~T. The observation of the higher-energy particles (\textbf{m}$_3$ to \textbf{m}$_6$) above the low-energy two-particle continua features quantum many-body effects in the exotic dynamic spectrum.
\end{abstract}

\maketitle

\section{Introduction}
\begin{figure}[t]
\centering
\includegraphics[width=\linewidth]{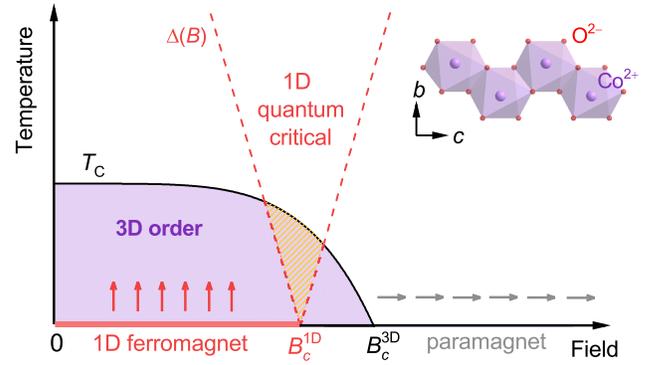}
\caption{Illustration of phase diagram for a quasi-1D ferromagnetic Ising-chain system in an applied transverse field. For a 1D ferromagnet, a long-range order is formed only at zero temperature, whereas a 3D order can be stabilized at a finite temperature $T_C$ in presence of interchain couplings. The 1D and 3D long-range orders can be suppressed by an applied transverse field at $B_c^{\text{1D}}$ and $B_c^{\text{3D}}$, respectively.
When $B_c^{\text{1D}}<B_c^{\text{3D}}$ the $E_8$ dynamic spectrum could be realized around $B_c^{\text{1D}}$ as illustrated by the dashed area. Inset shows the zig-zag spin chain constituted by edge-shared CoO$_6$ octahedra in CoNb$_2$O$_6$.}
\label{fig:PD}
\end{figure}

Since its invention in 1920 the Ising spin-chain model \cite{Lenz1920,Ising1925,Brush67} has been demonstrated to be extremely useful to rigorously illustrate basic concepts, thus the study of the Ising spin chains is still a very lively research field \cite{Sachdev,Dutta,Mussardo,Pfeuty70,McCoy78,Zamolodchikov89,Delfino95,
Delfino96,HUSSON1977,Hanawa94,Nojiri99,
Coldea10,Balents2010,Moore11,Armitage14,
Kinross14,Wu14,Cabrera14,Robinson14,Liang15,Wang15a,Wang16,
ZW18,Wang18a,Faure18,Wang19,Yang19,James19,Yu19,Bera20,Fava20,Zhang20,Zou20}. For example, a quantum phase transition occurs in the transverse-field Ising-chain model
\begin{equation}
    H = -J \sum_i S_i^zS_{i+1}^z-B\sum_i S_i^x
\end{equation}
when the spin gap $\Delta$ is closed at the critical field $B_c=J/2$ with $J$ being the exchange interaction between the nearest-neighbor spin-1/2 magnetic moments $\mathbf{S}_i$ on a chain (see Fig.~\ref{fig:PD}).
The transverse-field Ising-chain quantum critical point is characterized by a peculiar thermodynamic property: With decreasing temperature at the critical field, the Grüneisen parameter converges \cite{Wang19,Wu18,Zhang19}, in contrast to the divergent behaviour for a generic quantum critical point \cite{Zhu03}.

The quantum spin dynamics also exhibits exotic features close to this quantum critical point. When the transverse-field Ising chain is perturbed by a small longitudinal field $B_z$ via the Zeeman interaction $-B_z\sum_i S_i^z$, it was predicted that an exotic dynamic spectrum emerges around $B_c$, exhibiting eight particles with specific mass ratios (see Fig.~\ref{fig:E8}) \cite{Zamolodchikov89}. 
This exotic spectrum is uniquely described by a quantum integrable field theory with the symmetry of the $E_8$ Lie algebra \cite{Zamolodchikov89,Delfino95}.
(Introductory discussions of the $E_8$ Lie algebra in mathematics and in the relevant context of quantum field theory can be found in Refs.~\cite{Vogan07,Borthwick2010}.)
As this model cannot be represented by single-particle states but is featured by many-body interactions, it is challenging to find an exact analytical solution beyond the quantum critical point.

Despite its celebrity in mathematics \cite{Vogan07,Borthwick2010}, the $E_8$ symmetry has rarely been explored experimentally. Until 2010 the first piece of experimental evidence for the $E_8$ dynamic spectrum was reported based on inelastic neutron scattering measurements of the ferromagnetic Ising chains in CoNb$_2$O$_6$ \cite{Coldea10}. Constituted by edge-shared CoO$_6$ octahedra, the effective spin-1/2 chains in CoNb$_2$O$_6$ run along the crystallographic \textit{c} axis in a zig-zag manner (see inset of Fig.~\ref{fig:PD}), with the Ising easy axes lying in the crystallographic \textit{ac} plane \cite{HUSSON1977,Hanawa94,Liang15}.
Due to small but finite inter-chain couplings, a three-dimensional (3D) magnetic order develops below $T_C=2.85$~K, which can be suppressed by an applied transverse field of $B^{3D}_c \approx 5.3$~T along the \textit{b} axis (see Fig.~\ref{fig:PD} for an illustration) \cite{HUSSON1977,Hanawa94,Liang15,Matsuura20}.
By following the low-lying spin excitations in the transverse field ($B \parallel b$), two modes were found with an energy ratio being the Golden ratio $(1+\sqrt{5})/2 \approx 1.618$ at 5~T \cite{Coldea10}, which corresponds to the predicted mass ratio \textbf{m}$_2$/\textbf{m}$_1$ of the first two $E_8$ particles \cite{Zamolodchikov89}. 
The two excitations were interpreted as the low-lying $E_8$ particles, although the higher-energy $E_8$ particles were not resolved \cite{Coldea10}.
The required effective longitudinal field for realizing the $E_8$ dynamic spectrum is provided by interchain interactions in the ordered phase, and the corresponding one-dimensional (1D) quantum critical point at $B^{1D}_c \approx 5$~T \cite{Coldea10} is located below the 3D quantum phase transition at $B^{3D}_c \approx 5.3$~T (see Fig.~\ref{fig:PD}) \cite{Kinross14,Liang15}.

The absence of the higher-energy $E_8$ particles \textbf{m}$_3$ to \textbf{m}$_8$ in the inelastic neutron scattering spectra was assumed to a be consequence of an overwhelming (\textbf{m}$_1+$\textbf{m}$_1$) continuum \cite{Coldea10}, since it is energetically more favourable to excite two \textbf{m}$_1$ particles, as \textbf{m}$_3 \lesssim 2$\textbf{m}$_1 < $ \textbf{m}$_4$, \textbf{m}$_5$, …, \textbf{m}$_8$.
This notion may be more natural for non-interacting particles, but 
it is not necessarily applicable to the concerned quantum many-body system \cite{Moore11,Zhang20,Zou20}.
Using the time evolving block decimation, a numerical study \cite{Moore11} showed that the higher-energy $E_8$ particles up to \textbf{m}$_5$ should stand out as sharp peaks in the dynamic spectrum, whereas the (\textbf{m}$_1+$\textbf{m}$_1$) continuum contributes a relatively small background. 
Moreover, the two-particle continuum (\textbf{m}$_1+$\textbf{m}$_2$) was found to be characterized by a peak-like maximum at the onset energy, and thus potentially resolvable on top of the (\textbf{m}$_1+$\textbf{m}$_1$) continuum.

Very recently the findings of the numerical simulations were supported by rigorous quantum field-theory analysis of the dynamic spectra of the two-particle continua \cite{Zhang20,Zou20}, which revealed that the spectral weight of the (\textbf{m}$_1+$\textbf{m}$_1$) continuum decreases considerably with increasing energy, becoming relatively weak particularly at the energies where the higher-energy particles are predicted to appear. Moreover, it showed that the two-particle continua, such as (\textbf{m}$_1+$\textbf{m}$_1$) and (\textbf{m}$_1$+\textbf{m}$_2$), are not featureless but characterized by a peak-like maximum at the onset energies which is followed by an extended tail towards higher energies \cite{Zhang20,Zou20}.  These theoretical results clearly showed the exotic dynamic features of this quantum many-body system, in contrast to the conventional understanding drawing from a single-particle picture. Motivated by these theoretical results, we experimentally revisited the spin dynamic spectrum in CoNb$_2$O$_6$ by performing high-resolution terahertz spectroscopy in an applied transverse magnetic field. We identify not only the two lowest $E_8$ particles but also the higher-energy ones up to \textbf{m}$_6$, as well as the peak-like maximum of the two-particle continuum (\textbf{m}$_1$+\textbf{m}$_2$), confirming the theoretical predictions of the 1D quantum many-body system \cite{Moore11,Zhang20,Zou20}.

\section{Experimental details}

Single crystals of CoNb$_2$O$_6$ were grown by the floating-zone technique, following 
the procedure reported in Ref.~\cite{Prabhakaran03}, with few modifications. We used polycrystalline powders of Co$_3$O$_4$ (chemical purity 99.9985\%) and
Nb$_2$O$_5$ (99.9985\%) as starting materials. Two powder reactions were performed in air at 1200$^\circ$C
and 1250$^\circ$C, respectively, each for 12 h. The powder was pressed to a cylindrical rod
at 50 MPa, then sintered at 1275$^\circ$C. A centimeter-sized single crystal was grown in an
atmosphere of 80\% O$_2$/20\% Ar and small over-pressure with a growth speed of 3 mm/h
and a relative rotation of the rods of 30 rpm. X-ray powder diffraction measurements
verified phase purity. Laue images confirmed single crystallinity, and were used for cutting \textit{b}-axis oriented plate-like samples of about 3~mm in diameter and a thickness of 0.5~mm for the optical measurements.
On smaller samples magnetic susceptibility measurements were performed in a 100~mT field $B \parallel b$ down to 1.8 K confirming the magnetic transitions at 2.9~K and 1.9~K \cite{Hanawa94,Liang15}.

Using a Sciencetech SPS200 Martin-Puplett type spectrometer, field dependent THz transmission measurements were carried out at 4~K (above $T_C$) and 0.25~K (below $T_C$) with a
liquid-helium bath cryostat and a ${}^3$He-${}^4$He dilution fridge, respectively, using bolometers operating at 0.3~K and 0.4~K as detectors.
For the 4~K experiment, a rotatable polarizer was placed in front of the sample for tuning polarization of the THz waves \cite{Wang17,Zhang20}.
For the 0.25 K measurements the sample cell was attached to the cold finger of the dilution fridge (Oxford Instruments) which was equipped with a superconducting solenoid for applying a magnetic field.
The sample cell was filled with ${}^4$He gas at room temperature to provide cooling of the sample.
The radiation was filtered with a 0.6~THz low pass filter at 4\,K before the radiation entered the vacuum can of the dilution unit. A frequency resolution of 6~GHz was achieved in the measurements.
For the optical experiments, the THz radiation propagated in the direction of the external magnetic field which was applied parallel to the \textit{b}-axis of the CoNb$_2$O$_6$ single crystals.

\section{Experimental results and discussion}

\begin{figure}[t]
\centering
\includegraphics[width=0.9\linewidth]{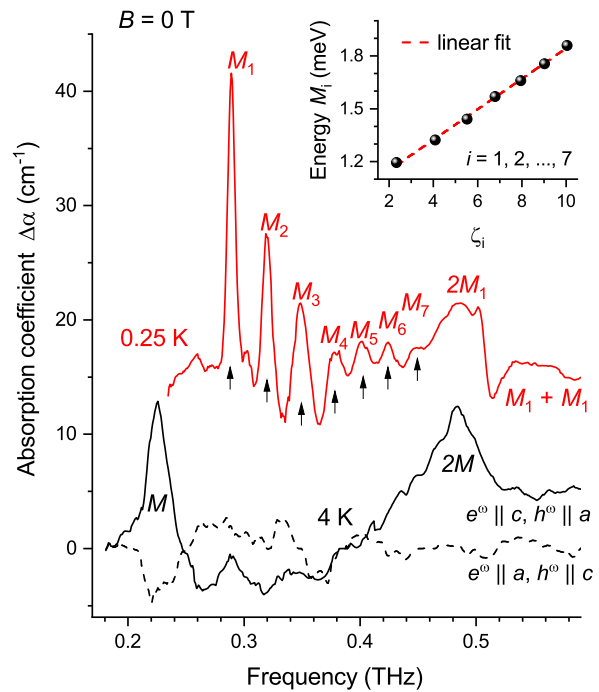}
\caption{Absorption spectra measured in zero field at 0.25~K (below $T_c$) with unpolarized THz radiation, and at 4~K (above $T_c$) for the THz polarizations ($e^\omega \parallel c$, $h^\omega \parallel a$) and ($e^\omega \parallel a$, $h^\omega \parallel c$). Inset shows that the eigenenergies of the modes $M_1$ to $M_7$ observed at 0.25~K follow a linear dependence on $\zeta_i$ which are the negative zeros of the Airy function $Ai(-\zeta_i)=0$. The linear dependence is expected for confined-spinon excitations. The different nomenclature for these zero-field excitations
is used in order to discriminate them from the $E_8$ particles close to the 1D criticial field $B_c^{1D}=4.75$~T (see below).}
\label{fig:0T}
\end{figure}

Zero-field absorption spectra are displayed in Fig.~\ref{fig:0T} for 0.25~K with unpolarized THz radiation, and for 4~K with the THz electric field $e^\omega \parallel c$ and the THz magnetic field $h^\omega \parallel a$, and with the polarization ($e^\omega \parallel a$, $h^\omega \parallel c$).
At 4~K the spectrum of ($e^\omega \parallel c$, $h^\omega \parallel a$) exhibits two peaks at 0.22 and 0.48~THz, respectively, which are denoted by $M$ and $2M$.
The nomenclature hereafter for the zero-field excitations is discriminated from that of the $E_8$ particles. Around the $2M$ peak one can observe a broad continuum-like feature which extends towards higher frequency.
These features are similar to those reported for a different polarization ($e^\omega \parallel a$, $h^\omega \parallel b$) in Ref.~\cite{Armitage14}, where the $M$ and $2M$ peaks were assigned as the one- and two-pair spinon excitations, respectively.
In contrast, these features are absent for the polarization ($e^\omega \parallel a$, $h^\omega \parallel c$) (see Fig.~\ref{fig:0T}).

Compared with the 4~K spectrum, the 0.25~K one below $T_C$ exhibits more peaks, which are labeled by $M_i$ ($i=1,2,3,...,7$) with increasing frequency. The energies of $M_i$ are shown in the inset of Fig.~\ref{fig:0T} as a function of $\zeta_i$, the negative zeros of the Airy function $Ai(-\zeta_i)=0$. The linear dependence on $\zeta_i$ reflects the spinon confinement in a linear confining potential \cite{McCoy78,James19}, which is set up by the inter-chain couplings in the magnetically ordered phase \cite{Coldea10,Armitage14,Wang15a,Wang18a,Faure18,Wang19,Zhang20}.
Above $M_7$ one can see a broader peak around 0.49~THz (labelled $2M_1$) and a broad continuum at higher energy (labelled $M_1+M_1$), consistent with the observation in Ref.~\cite{Armitage14}. The $2M_1$ peak corresponds to a kinetic bound state of two pairs of spinons in neighboring chains, which is located below the excitation continuum of two-independent pairs of spinons ($M_1+M_1$) \citep{Coldea10,Armitage14}.
This bound state was found at the Brillouin zone boundary $q=\pi$ by inelastic neutron scattering \cite{Coldea10}.
Due to the zig-zag configuration of the chains (see inset of Fig.~\ref{fig:PD}) \cite{Fava20}, this mode is folded to the zone center ($q=0$) and thus detected by the THz spectroscopy. 
We shall emphasize that the spinon dynamics and the $E_8$ spectrum are about very different physics. The former is about the spin dynamics of the gapped phase at zero field, whereas the latter emerges only around the field-induced quantum critical point.

\begin{figure}[t]
\centering
\includegraphics[width=\linewidth]{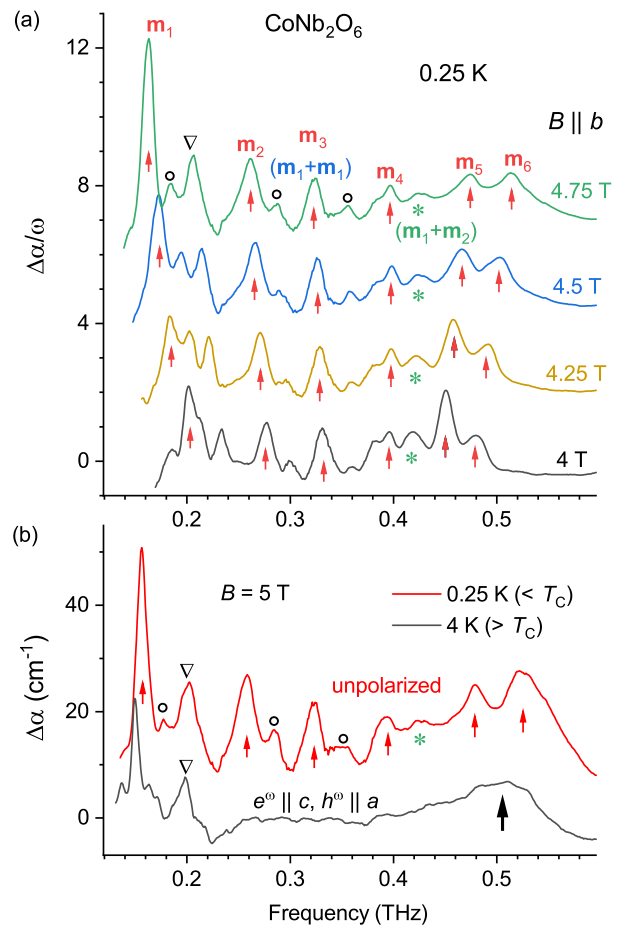}
\caption{(a) Absorption spectra of CoNb$_2$O$_6$ measured at 0.25~K in various applied transverse magnetic fields, $B \parallel b$. $\Delta \alpha$ and $\omega$ denote absorption coefficient and wavenumber, respectively, in the unit of cm$^{-1}$. The arrows indicate the modes \textbf{m}$_1$, \textbf{m}$_2$,..., \textbf{m}$_6$ at 4.75~T and their field-dependent evolution. The asterisk "$\ast$" marks the onset of the (\textbf{m}$_1$+\textbf{m}$_2$) continuum. At 4.75~T the circles "$\circ$" mark the satellite peaks. The spectra in higher fields are shifted upward by a constant for clarity. (b) Absorption spectra measured at 5~T for 0.25 and 4~K below and above $T_C$, respectively. For 0.25~K the arrows indicate those modes marked by arrows in (a). The black arrow at the 4~K spectrum marks a broader band observed due to the zone-folding effect. The triangle "$\triangledown$" indicates a mode which similar to $m_1$ is present both above and below $T_C$. The 0.25~K spectrum is shifted upward for clarity.}
\label{fig:spectra}
\end{figure}

An ideal way to study the $E_8$ dynamic spectrum would be to first drive an Ising-chain system with an applied transverse field to the quantum critical point, and then monitor the evaluation of the spin dynamics by switching on and tuning a perturbative longitudinal field.
However, such tuning can hardly be realized in a solid-state material, where an effective longitudinal field is an internal field determined by the inter-chain couplings.
Since the transverse field will compete with the inter-chain couplings, the 1D quantum critical point may not be reached before the 3D order is suppressed (e.g. in SrCo$_2$V$_2$O$_8$ \cite{Wang15a,Yu19}). 
To realize the $E_8$ spectrum, the 1D quantum critical point should be hidden in the 3D ordered phase as illustrated in Fig.~\ref{fig:PD}, which is fulfilled in the Ising-chain ferromagnet CoNb$_2$O$_6$ \cite{Coldea10,Kinross14} and in the Ising-chain antiferromagnet BaCo$_2$V$_2$O$_8$ \cite{Faure18,Zhang20,Zou20}.
This also indicates that the observation of a spinon confinement in zero field does not necessarily imply a realization of the $E_8$ dynamic spectrum around the quantum critical field. Therefore, it is necessary to carry out field-dependent measurements below $T_c$.
 
The evolution of the absorption spectra of CoNb$_2$O$_6$ in an applied transverse field along the $b$ axis is presented in Fig.~\ref{fig:spectra} for fields just below 5~T, at which the inelastic neutron scattering experiment \cite{Coldea10} revealed the lowest two $E_8$ particles \textbf{m}$_1$ and \textbf{m}$_2$. With far more than two peaks, the absorption spectra exhibit very rich features. At 4.75~T one observes several well-defined sharp peaks at 0.16, 0.26, 0.32, 0.40, 0.47, and 0.51~THz, which are labelled \textbf{m}$_1$, \textbf{m}$_2$, ... \textbf{m}$_6$, respectively, as indicated by the arrows. A relatively broad peak is observed at 0.43~THz as marked by the asterisk.
The frequencies of \textbf{m}$_1$ and \textbf{m}$_2$ are slightly greater than the reported values of 0.12 and 0.18~THz, respectively, for the finite \textit{q}-vector $(3.6,0,0)$ by the inelastic neutron scattering experiment \cite{Coldea10}. This difference may result from a weak dispersion perpendicular to the chain direction.

\begin{figure}[t]
\centering
\includegraphics[width=\linewidth]{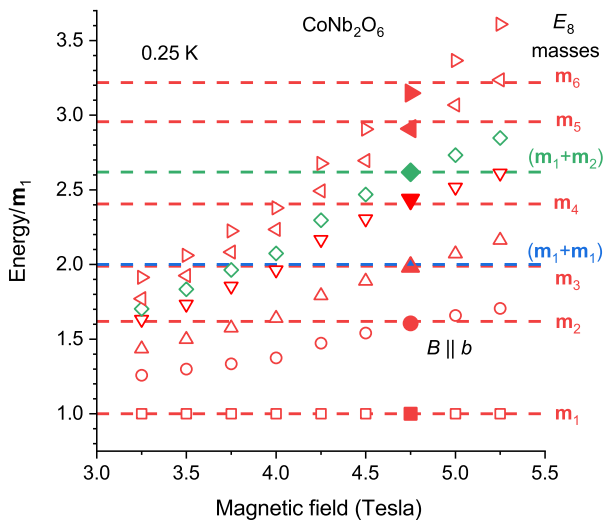}
\caption{Field dependence of the energies of the observed modes \textbf{m}$_1$, ... \textbf{m}$_6$ and of the maxima for (\textbf{m}$_1$+\textbf{m}$_1$) and (\textbf{m}$_1$+\textbf{m}$_2$), normalized to the \textbf{m}$_1$ energy in each field. All the ratios exhibit a monotonic increase with increasing field. The predicted ratios (dashed lines, see Ref.~\cite{Zamolodchikov89,Delfino95}) for the $E_8$ particles are simultaneously reached at $B^{\text{1D}}_c = 4.75$~T, evidencing the observation of the high-energy $E_8$ particles in CoNb$_2$O$_6$.}
\label{fig:E8}
\end{figure}

The field dependence of these modes can be clearly tracked, as indicated by the arrows in Fig.~\ref{fig:spectra}(a)(b). Normalized to the \textbf{m}$_1$ energy in each field, the eigenenergies of these modes are presented as a function of the applied field in Fig.~\ref{fig:E8}. The energy ratios of these modes increase monotonically with increasing field.
At 4.75~T the predicted ratios (dashed lines, see Refs.~\cite{Zamolodchikov89,Delfino95}) for the $E_8$ particles up to \textbf{m}$_6$ and for the onset energies of the two-particle continua (\textbf{m}$_1+$\textbf{m}$_1$) and (\textbf{m}$_1$+\textbf{m}$_2$) are simultaneously reached, evidencing the observation of the $E_8$ dynamic spectrum.
The onset of the (\textbf{m}$_1+$\textbf{m}$_1$) continuum is very close to the \textbf{m}$_3$ peak ($\approx$~1.989\,\textbf{m}$_1$) \cite{Zamolodchikov89}, so they cannot be distinguished from each other in the experimental spectrum.
The observed features are consistent with the previous predictions from the numerical simulations \cite{Moore11} and the quantum field-theory analysis \cite{Zhang20,Zou20}. Moreover, the field-theory analysis \cite{Zhang20,Zou20} showed that the two-particle continua are not featureless but characterized by a peak-like maximum at the onset energies followed by a continuous decrease of spectral weight towards higher energy, which allows the identification of the continua by their peak-like maxima.
Therefore, these experimental results provide unambiguous evidence for the observation of the high-energy $E_8$ particles, which also points to a hidden 1D quantum critical point at $B^{1D}_c = 4.75$~T confirming the scenario illustrated in Fig.~\ref{fig:PD} and discussed above. The value of $B^{1D}_c$ is close to the reported 5~T in Ref.~\cite{Coldea10}.

Previous theoretical analysis also predicted that the intensity of the $E_8$ particles decreases monotonically with increasing energy \cite{Delfino95,Delfino96,Moore11,Zhang20,Zou20}. Indeed, this trend is obeyed by the first four particles (\textbf{m}$_1$ to \textbf{m}$_4$), as shown in the 4.75~T spectrum in Fig.~\ref{fig:spectra}(a). 
However, the \textbf{m}$_5$ and \textbf{m}$_6$ peaks appear to be slightly stronger. This cannot be simply attributed to the underlying continua (\textbf{m}$_1$+\textbf{m}$_3$) or (\textbf{m}$_2+$\textbf{m}$_2$) whose spectral weight is even smaller than the high-energy tails of the (\textbf{m}$_1+$\textbf{m}$_1$) and (\textbf{m}$_1$+\textbf{m}$_2$) continua \cite{Moore11,Zhang20,Zou20}.
The apparent enhancement of the \textbf{m}$_5$ and \textbf{m}$_6$ peaks is contributed by the low-lying spin excitation at the Brillouin-zone boundary ($q=\pi$)  \cite{Cabrera14,Robinson14}. This relatively broad band is detected also in the disordered phase above $T_c$, as indicated by the arrow in the 4~K spectrum in Fig.~\ref{fig:spectra}(b), which is observed due to the zone-folding effects \citep{Fava20}. It is a coincidence that this band is located in the energy range around the \textbf{m}$_5$ and \textbf{m}$_6$ peaks. 
The substantially reduced intensity of the high-energy $E_8$ particles could be below the resolution limit of the previous inelastic neutron scattering experiment \cite{Coldea10}, which thus were not resolved at that time.
For the same reason the \textbf{m}$_7$ and \textbf{m}$_8$ modes are not resolved here either.

The field dependence of the relatively small satellite peaks, marked by the circles in Fig.~\ref{fig:spectra}(a), can be clearly followed as well. With decreasing field from 4.75~T one can see a reduction of the satellite-peak intensity and a concomitant merging of these peaks into the corresponding main ones.
Above $T_c$ in the disordered phase [Fig.~\ref{fig:spectra}(b)], these satellite peaks disappear, thus they reflect dynamic properties of the 3D ordered phase in the transverse field.
In addition, as marked by the triangles in Fig.~\ref{fig:spectra}(b), one can observe a peak at 0.2~THz both in the ordered and in the disordered phases. Thus, this mode should result from the 1D spin fluctuations possibly a zone-boundary excitation observed due to sub-leading interactions within the zig-zag chain \cite{Fava20}.

To conclude, by performing high-resolution THz spectroscopy of the Ising-chain compound CoNb$_2$O$_6$ below and above the magnetic ordering temperature in an applied transverse field, we have revealed the dynamic features that were predicted to emerge around the transverse field-induced quantum critical point governed by the $E_8$ symmetry.
In particular, the high-energy $E_8$ particles, which would be unresolvable according the picture of non-interaction particles, have been identified above the low-energy two-particle continua, featuring the quantum many-body effects. We have also observed features beyond the $E_8$ dynamics, which appeals for a theoretical study of a realistic model for CoNb$_2$O$_6$.

\begin{acknowledgments}
We thank Jianda Wu and Zhao Zhang for insightful discussions, and Thomas Timusk for the help in constructing the bolometer unit for the dilution fridge.
The work in Tallinn was supported by personal research funding grant PRG736 of the Estonian Ministry of Education and Research, and by European Regional Development Fund Project No. TK134. The work in Cologne was partially supported by the DFG (German Research Foundation) via the project No. 277146847—Collaborative Research Center 1238: Control and Dynamics of Quantum Materials (Subprojects No. A02, B01, and B05).
\end{acknowledgments}

\bibliographystyle{apsrev4-1}
\bibliography{CNO_bib}

\end{document}